\documentclass[%
 reprint,
 superscriptaddress,
 amsmath,amssymb,
 aps,
]{revtex4-2}

\usepackage{multirow}
\usepackage{graphicx}
\usepackage{dcolumn}
\usepackage{bm}
\usepackage{hyperref}
\usepackage{textcomp}

\hypersetup{
    colorlinks=true,       
    linkcolor=blue,        
    citecolor=blue,        
    filecolor=magenta,     
    urlcolor=blue          
}

\begin{document}


\title{The Lowest Broad Alpha Cluster Resonances in $^{19}$F}

\author{A. Volya}%
\email{volya@phy.fsu.edu}
\affiliation{Department of Physics, Florida State University, Tallahassee, 32306-4350, USA}%
\affiliation{Cyclotron Institute, Texas A\&M University, College Station, Texas, 77843-3366, USA}%
\author{V. Z. Goldberg}%
\affiliation{Cyclotron Institute, Texas A\&M University, College Station, Texas, 77843-3366, USA}%

\author{A. K. Nurmukhanbetova}%
 \affiliation{Energetic Cosmos Laboratory, Nazarbayev University, Nur-Sultan, 010000, Kazakhstan}%
 
\author{D. K. Nauruzbayev}%
\affiliation{Nazarbayev University Research and Innovation System, Nur-Sultan, 010000, Kazakhstan}%
\affiliation{Saint Petersburg State University, Saint Petersburg, 199034, Russia}%

 \author{G. V. Rogachev}
 \affiliation{Cyclotron Institute, Texas A\&M University, College Station, Texas, 77843-3366, USA}%
 \affiliation{Department of Physics and Astronomy, Texas A\&M University, College Station, Texas 77843, USA}%
 \affiliation{Nuclear Solutions Institute, Texas A\&M University, College Station, Texas, 77843, USA}%
 
\date{\today}

\begin{abstract}
\begin{description}
\item[Background]
There is   a deep astrophysical interest in the structure of $^{19}$F states close to the alpha decay threshold.  The nuclear structure of these states is important for understanding { of the} development of $\alpha$ clustering in the $^{20}$Ne region. 
{Emergence of clustered states and generally states that favor coupling to reaction channels near the corresponding decay thresholds is currently of special interest in theoretical physics.}
The only  detailed high energy resolution $^{15}$N($\alpha$,$\alpha$) work was made in 1961, and a complete analysis of these data has never been made.
\item[Purpose]
To identify the parameters of broad low spin states in $^{19}$F  near the $\alpha$ decay threshold,  {to present a theoretical study of these states and to assess the current state of the theoretical methods.}
\item[Method]
Excitation function for $^{15}$N($\alpha$,$\alpha$) elastic scattering was measured by the TTIK method. These new data together with  old, high energy resolution data,  were analyzed using the $R$ matrix approach. {$^{19}$F nuclear structure was calculated using configuration interaction methods with the recently developed effective interaction Hamiltonian.}

\item[Results] 
The parameters of broad low spin $\ell=0$ and 1 relative partial wave resonances close to the $\alpha$ decay threshold in $^{19}$F were {identified.  Detailed theoretical analysis was carried out identifying all states coupled to the $\ell=0$ and 1 alpha cluster channels. 
Considering hierarchy of states with different harmonic oscillator shell excitations allows to evaluate coupling to the alpha channels with different 
number of nodes in the relative wave function and helps to explain the distribution of the clustering strength and emergence of broad clustering resonances. Comparison of clustering in 
$^{20}$Ne into $^{16}$O+$\alpha$ and consideration of spin-orbit splitting of the $^{15}$N+$\alpha$ channel provides additional evidence. } 

\item[Conclusions] {Detailed analysis of new and old experimental data allows to identify a series of $\alpha$ clustering resonances 
in $^{19}$F and to assess the distribution of the clustering strength which is of importance to questions of astrophysics and for theoretical understanding of many-body physics and emergence of clustering in loosely bound or unstable nuclei. Progress has been made in theoretical
understanding of the origins of clustering and questions for future theoretical and experimental research are identified. }

\end{description}
\end{abstract}
\maketitle


\section{\label{sec:level1}{Introduction}}

Fluorine is an element with an uncertain and widely debated cosmic origin. It has only one stable isotope, $^{19}$F, whose production and destruction is directly connected to the physical conditions in stars \cite{01}. 

Asymptotic giant branch (AGB) stars, {where $^{19}$F has been found via observations, }are considered as an important source of $^{19}$F in the Galactic [see Ref. \cite{02}, and references therein]. In AGB stars fluorine can be produced via {reactions} $^{14}$N$(\alpha, \gamma)$$^{18}{\rm F}(\beta^+),$  $^{18}$O$($p$, \alpha),$ and  $^{15}$N$(\alpha, \gamma)$$^{19}{\rm F}.$

Nuclear structure of $^{19}$F might be important for understanding production of the long-lived radioisotope $^{18}$F in novae and {in} heavy element production in x-ray bursts \cite{03,04,05,fortune:2020,fortune:2006}. {Here, the important reactions are} $^{18}$F$(p, \alpha)^{15}$O  
and  $^{18}$F$(p, \gamma)^{19}$Ne . These reactions proceed through the $^{19}$Ne nucleus. 
The authors \cite{03,04} noted that the needed information can be more {easily} obtained through studies of $^{19}$F, mirror to $^{19}$Ne nucleus.

{Interest} to the $^{19}$F nucleus is also supported by {the} general interest to the $\alpha$ cluster structure in atomic nuclei, well known in nearby $^{20}$Ne \cite{06}. Recently it was shown  \cite{06,07} that the $\alpha$ cluster structure in odd-even $^{21}$Ne nucleus has  striking similarities to that of 4N nucleus $^{20}$Ne. {$^{19}$F  is important for comparison of the structures 
$(^{17}{\rm O}+\alpha)$ and $(^{15}{\rm N}+\alpha)$ where  $^{17}{\rm O}$ has an extra nucleon and $^{15}{\rm N}$ has a hole relative to 
$^{16}{\rm O}$ core.}
{The theoretical interest towards understanding quantum many-body physics on the verge of stability has been rapidly expanding 
fueled by advances in experimental physics and emerging opportunities to test some of the theoretical models such as those found in
Refs.~\cite{volya:2009a:art,volya:2006a:art,michel:2003}, remarkable predictions such as superradiance \cite{auerbach:2011}, near threshold 
dynamics \cite{michel:2007a} and exotic virtual transitions \cite{volya:2015aa:proc}. Alpha clustering along with many other examples 
\cite{okolowicz:2020} represents a curious manifestation of the near-threshold resonances with significant collectivization of spectroscopic strength towards corresponding channels; this phenomenon is not fully understood.}

{In our previous works \cite{06,07} we made first steps towards explaining the properties of $\alpha$ cluster states in $^{20, \,21}$Ne using the recently developed configuration interaction methods for clustering \cite{08,09}. }
These application of these theoretical models showed that description of the low laying cluster states with positive parity was quite reasonable, however, {the approach did not reproduce low-lying clustering strength in the negative parity cluster states and some very broad alpha resonances were not present in the theory at all.  These results motivate the research work reported here by outlining the direction  in which the theoretical approach must be 
improved.} These theoretical developments are important both for better understanding of the clustering in atomic nuclei and for calculation of nuclear reaction induced by alpha particles in stars. Indeed, one has to realize  that many nuclear reactions important for astrophysics cannot be tested in laboratories because of desperately small cross sections.

In this work, we present new data on the lowest broad resonances in $^{19}$F, the properties of which were point of the astrophysical interest in  Ref. \cite{03,04,05} and an analysis of the properties of $\alpha$ cluster states in $^{19}$F in the framework of the recently developed theoretical methods \cite{08,09,10}.

\section{Experimental Data}

{The only high energy resolution, broad energy and angle  range  experimental study} of resonances in the $\alpha+^{15}$N  scattering was published over a half a century ago {in Ref. \cite{11}. The study covers a }broad interval of the excitation energies in $^{19}$F from 5.37 to 8.33 MeV with energy resolution about 0.1\% $E.$  All of the angles at which the excitation functions were measured in Ref. \cite{11}, except $169.1^\circ$, correspond to zeros of Legendre polynomials. The parameters of the observed resonances below 3.7 MeV of $\alpha$ particle energy were obtained  using phase shift analysis.  The  first  $R$ matrix analysis  of  data \cite{11} was performed T. Mo and H. Weller  \cite{12}  in 1972. This analysis resulted in a fair description of the data \cite{11} for  {$E_\alpha$} 3.8 to 4.8 MeV and yielded spin and parity assignments for five levels in $^{19}$F. A contemporary $R$ matrix analysis was made by authors of Ref. \cite{03}. The authors {of the work} \cite{03} were mainly motivated by the need for a more precise knowledge of the parameters {of broad low spin resonances needed for calculations of the reaction rates in astrophysics. } This analysis   \cite{03}   also corrected multiple errors in the previous spins assignments for the levels with $J=\ell\pm1/2$ which can be populated with the same  orbital {angular} momentum of captured   $\alpha$ particle. Unfortunately, the authors {of Ref.} \cite{03} could perform the analysis { of the available digital data from Ref. \cite{11} only for the $169.1^\circ$ angle. }

\begin{table*}[!t]

\caption{\textsuperscript{19}F levels\label{tablL0}}
\begin{center}
\begin{tabular}{|r|c|ll|ll|ll|ll|ll|ll|}
\hline
{N}&{J$^\pi$}&{Ref. \cite{17}}&&{Ref.\cite{03}}&&{Ref.\cite{11}}&&{Ref.\cite{04}}&&{Ref.\cite{05}}\mbox&&\mbox{}{This work}&\\
\cline{3-4}\cline{5-6}\cline{7-8}\cline{9-10}\cline{11-12}\cline{13-14}
 & &E$'\footnote{Excitation energy in $^{19}$F}$&$\Gamma_{\alpha}\footnote{alpha width}$&E$_x$&$\Gamma_\alpha$&E$'$&$\Gamma_\alpha$&E$'$&$\Gamma_\alpha$&E$'$&$\Gamma_\alpha$& E$'$&$\Gamma_\alpha$\\
 &&(MeV)&(keV)&(MeV)&(keV)&(MeV)&(keV) &(MeV)&(keV)&(MeV)&(keV)&(MeV)&(keV)\\
\hline

1&$1/2^+$ &- &- & - & - &- &- & 5.337 & 1.3±0.5 & 5.336 & 2.51±0.10& 5.333&1.4±0.4\\

2&$3/2^+$&5.501&4±1&5.496&3.2&5.475&4&5.501&4.7±1.6&5.501&6.0±0.3&5.488&4.85±0.5\\

3&$5/2^+$&6.282±2&2.4&6.289&2.4&6.269&3 &- &- &- &-&6.289&2.30±0.5\\

4&$7/2^+$&6.330±2&2.4&6.338&3.6±0.4&6.317&3 &- &- &- &-&6.339&3.30±0.4\\

5&$1/2^-$&6.429±8&280&6.536&245±6&6.41&358 &- &- &- &-&6.540&220±40\\

6&$1/2^-$&6.989±3&51&7.028&96±6&6.97&64&- &- &- &-&7.048&150±35\\
\hline
\end{tabular}
\end{center} 
\end{table*}

\begin{figure}[!t]
    \begin{center}
    \includegraphics[width=85mm]{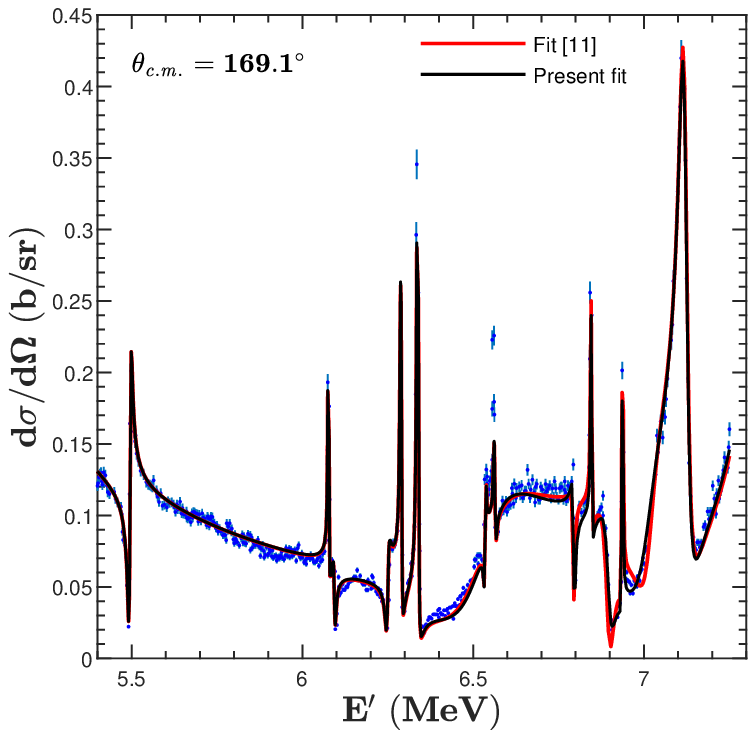}
    \includegraphics[width=85mm]{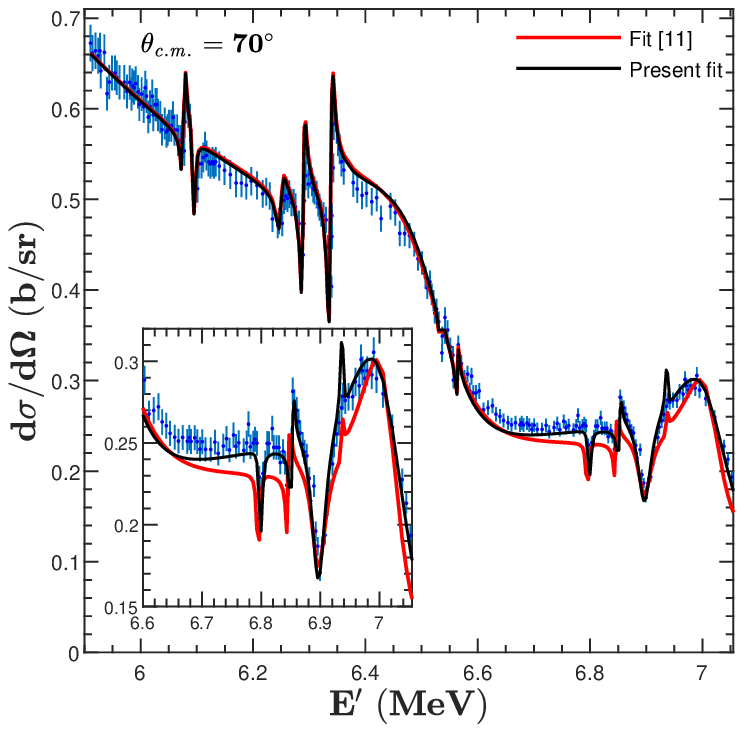}
      \end{center}
    \caption{\label{fig1_a.eps} 
       (a) $R$ matrix fit of the excitation functions for the $\alpha+^{15}$N elastic scattering \cite{11} at  $169.1^\circ$ in comparison with the fit of the Ref.\cite{03}; (b) the same for $70^\circ$. An {inset  highlights the difference between the present fit} and the fit with parameters of Ref.\cite{03}.}

\end{figure}

More recently two experiments \cite{04} and \cite{05} were performed to obtain the total widths and the partial gamma decay widths for $1/2^+$ and $3/2^+$ states  at the 5.3 and 5.5 MeV excitation energy in $^{19}$F. 
However, the results \cite{04} and  \cite{05} on the widths of the states disagreed with each other {well beyond} the quoted uncertainties  (see Table \ref{tablL0}).

We have reconsidered the data on low spin broad resonances close to the $\alpha$ particle decay threshold in $^{19}$F. First, we reanalyzed the data \cite{11} again. {In difference} to Ref. \cite{03}   we  included in the analysis all data \cite{11}, at all measured angles, and in complete energy region; we also used a convolution of the $R$ matrix calculations with experimental energy resolution.

{In addition to the analysis of the existent data,} we performed measurements of the excitation functions for the $\alpha+^{15}$N elastic scattering by the Thick Target Inverse Kinematic (TTIK) method \cite{13,14} to obtain new experimental data on the low lying and broad resonances in question.

{Recently a new detailed study of the $\alpha+^{15}$N elastic scattering with the TTIK method at higher excitation energy than in the present work was reported in Ref. \cite{15}.}


While the energy resolution in the TTIK method is worse than in the classical approach of Ref. \cite{11},  the observation of  the excitation functions at $180^\circ$ {which is at the } minimum of the dominant Rutherford scattering, and a better counting statistics {compensate
for the deficiencies of the TTIK approach used.}

\begin{figure}[!t]
    \begin{center}
    \includegraphics[width=85mm]{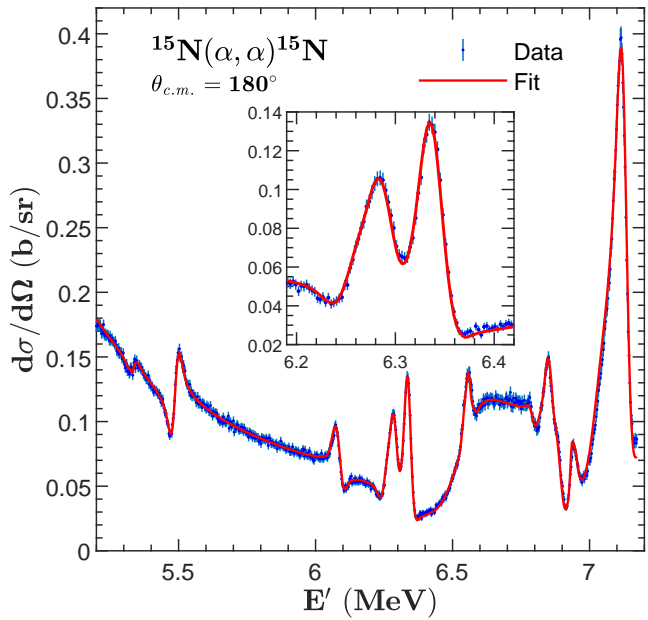}
    \end{center}
    \caption{\label{fig2.eps} 
    The excitation function for $\alpha+^{15}$N elastic scattering. {inset} caption: $R$ matrix fit of the excitation function for $\alpha+^{15}$N elastic scattering in the region of $5/2^+$ and $7/2^+$ resonances, using convolution with experimental energy resolution of 33 keV.}
\end{figure}

\begin{figure}[!t]
    \begin{center}
     \includegraphics[width=85mm]{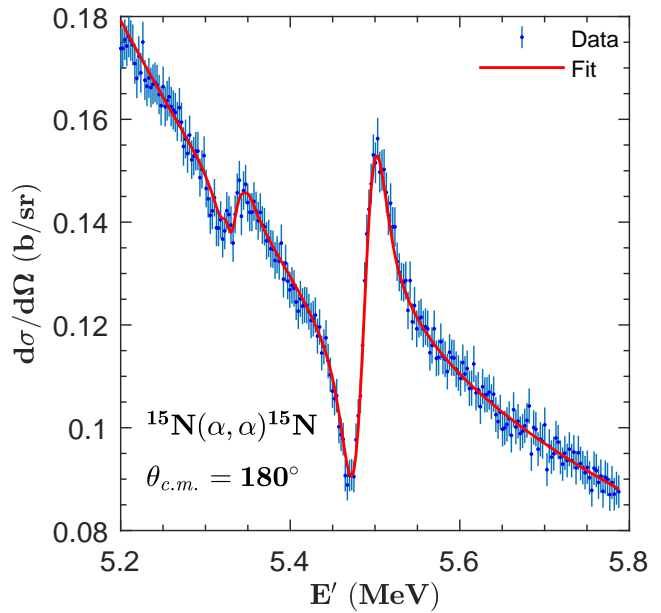}
     \end{center}
    \caption{\label{fig3.eps} $R$ matrix fit of the region of $1/2^+$ (5.3 MeV) and $3/2^+$ (5.5 MeV) levels in $^{19}$F measured  by TTIK method. The level parameters are given in Table \ref{tablL0}.}
    
\end{figure}

Fig.\ref{fig1_a.eps} shows an $R$ matrix fit of data of Ref.  \cite{11} at $169.1^\circ$. The $R$ matrix calculations were performed with the code AZURE \cite{16}.  We obtained a reasonable fit to the data \cite{11} at all angles and in the whole energy region of measurements \cite{11} up to 8.3 MeV excitation energy.  The full results of this analysis including over 50 resonances  will be published elsewhere. 

Fig.\ref{fig1_a.eps} was obtained using parameters for low spin states {that are different from those in Ref. \cite{03}, see Table \ref{tablL0}, however at  $169.1^\circ$ the  difference is hardly noticeable.  }
Usually, it is considered that the parameters of an $R$ matrix analysis are most sensitive to  the excitation functions measured at angles very close to $180^\circ$. This is because the potential scattering contribution decreases towards $180^\circ$, and the resonances are at their maximum. However, the $\ell$=0 resonances have an isotropic angular distribution, and can manifest {themselves}  as broad singularities at angles close to $90^\circ,$ while  the higher $\ell$ resonances {are} weaker.   Besides, it is worthwhile to note that the odd $\ell$ resonances should be very weak in $\alpha+^{15}$N resonant scattering at $90^\circ$; 
{the odd Legendre polynomials that describe the scattering of spinless ions are equal to zero at this angle.}
We  found that the fit with the parameters \cite{03}  deteriorates at angles close to $70^\circ$.  As it seen in Fig.\ref{fig1_a.eps}, the modified parameters for $\ell$=0 resonances (Table \ref{tablL0}) provide a better description of the experimental data. 
Fig.\ref{fig2.eps} demonstrates an excitation function for $\alpha+^{15}$N elastic scattering at $180^\circ$ obtained using the TTIK approach. The measurements were made at DC-60 facilities at Nur-Sultan (Kazakhstan) at $^{15}$N  beam energy of 21 MeV,  and all conditions were very similar to those described in Ref. \cite{13}. In the present work, the  excitation functions were used to obtain the parameters  of the $1/2^+$ and $3/2^+$ states at the 5.3 and 5.5 MeV excitation energy in $^{19}$F. A good knowledge of the experimental energy resolution is important because of the narrow width of the resonances. The energy resolution of the TTIK approach depends upon several parameters of the experiment, mainly upon energy of the beam, energy loss and straggling of the beam in the gas, gas pressure and energy resolution of the detectors. These parameters are known with different uncertainties. To obtain a reliable evaluation of the resolution  we made a fit of narrow $5/2^+,$ $7/2^+$ resonances in the TTIK  using data \cite{11}. An {inset} in Fig.\ref{fig2.eps} shows the excitation function in the region of the $5/2^+$ ($E'$=6.299 MeV), $7/2^+$ ($E'$=6.339 MeV) resonances using convolution with experimental resolution of 33 keV. {
Here and in tables $E'$ refers to the energy above the alpha particle separation energy.}
Then we used the obtained  energy resolution to fit the excitation function in the region of $1/2^+$ and $3/2^+$ resonances (Fig.\ref{fig3.eps}). To test a dependence of the influence of the energy resolution on the evaluation of the widths $1/2^+$ and $3/2^+$ resonances, we varied it by 10 \%. It resulted in ~0.1 keV variation of the widths.

Table \ref{tablL0} summarizes the known results for the $1/2^-,$ $1/2^+,$ and $3/2^+$  resonances in question. 
Taking into account the uncertainties of the available data the total widths are: $1/2^+$ ($E'$=5.33 MeV),  $\Gamma=1.4\pm 0.4$ keV and $3/2^+$ ($E'$=5.50 MeV);  $\Gamma=5.4\pm 0.4$ keV. 
Table \ref{tablL1} presents the reduced alpha particle widths of these states in $^{19}$F in comparison with the widths of the states with similar structure in $^{20}$Ne \cite{06}. It is evident from Table~\ref{tablL1} that the states with a similar core ($^{15}$N or $^{16}$O)+ $\alpha$ particle structure appear at the energies close to the alpha particle decay thresholds in $^{19}$F and $^{20}$Ne.  
We present theoretical consideration of the structure and clustering of these states in the next section. 

\begin{table}[htb]
\caption{$\alpha$ particle widths of the states with similar structure in $^{19}$F and $^{20}$Ne \cite{07}\label{tablL1}.}
\begin{center}
\begin{tabular}{|cccc|cccc|}
\hline
&{$^{20}$Ne}&&&&{$^{19}$F}&&\\
\cline{1-4}\cline{5-8}{J$^\pi$}&E$'$\footnote{E$'$ is the excitation energy relative to the alpha particle decay threshold (4.7 MeV for $^{20}$Ne and 4.0 MeV for $^{19}$F)}&$\Gamma_\alpha$ & $\gamma_\alpha$\footnote{reduced alpha width of the level}& {J$^\pi$}&E$'$&$\Gamma_\alpha$&$\gamma_\alpha$\\
&(MeV)&(keV)&&&(MeV)&(keV)&\\
\hline
$1^-$&1.1&0.028&1.4& $1/2^+$ &1.3&1.4&0.65\\
-&-&-&-& $3/2^+$&1.5&5.4&0.88\\
$0^+$&2.0&19&0.47&$1/2^-$&2.4&220&0.51\\
$0^+$&2.5&3.4&0.17&	$1/2^-$&3.0&150&0.12\\
\hline
\end{tabular}
\end{center}
\end{table}

\section{Theory}
Recent decades saw some significant advances in microscopic understanding of clustering phenomena that stem from ideas of quantum configuration mixing involving shell-model like wave functions with reaction channels. Building up on the ideas of the Resonating Group Method \cite{18,19,20,21}, its algebraic extensions, and related Generator Coordinate Methods \cite{22}
the new combined No-Core Shell Model with Resonating Group Method 
\cite{23,24,25} have been gaining a foothold in modern studies. 
A recent development of the cluster center-of-mass boosting technique \cite{26,09} has been a breakthrough method allowing to extend clustering studies to a much broader scope of nuclei \cite{08}.
The manipulation with the center-of-mass coordinate of each cluster wave function within common harmonic oscillator basis 
allows to built cluster reaction channels in a systematic way, retaining antisymmetry with respect to all nucleons, allowing for arbitrary types and arbitrary number of clusters, while departing from the restrictive algebraic techniques \cite{27} that become problematic when nuclei of very different sizes are involved and clusters are represented by complex configurations of many components.  
The ability to systematically expand  clustering studies from light nuclei treated by the no-core shell model using nucleon-nucleon interactions coming from fundamental principles to a more traditional shell model applications for heavy nuclei that rely on the effective interactions is important. This opens doors for practical studies of clustering using experimental data and helps in understanding how clustering emerges in the many-body physics, what are the effective cluster degrees of freedom and what are those generic properties of the many-body structure and reactions that allow for the clustered states to survive and coexist among many other states in a complex quantum many-body dynamics.

Despite the success of the many-body techniques that stem from ab-inito principles full microscopic description of spectra of sufficiently large nuclei is difficult unless effective interactions are considered which are at the core of the traditional shell model \cite{28}. 
The nuclear interaction Hamiltonian with parameters taken from fundamental principles but tuned by observations are widely used in practice. These Hamiltonians are remarkably accurate, some good examples describe thousands of states seen in experiments along with their properties; they reliably predict states and are often used for guidance in experimental searches. 
Recently a new FSU interaction Hamiltonian has been developed \cite{29,30,31} where cross-shell matrix elements between $p,$ $sd,$ and $fp$ shells have been determined using the latest data on masses and energies of intruder states. The FSU interaction Hamiltonian is among the most broad in its region of applicability covering a valence space  from the $s$ shell to $fp$ shell, it has been demonstrated to be remarkably accurate, and works well for exotic states with multi-particle  cross shell excitations that were not a part of the fit. 
The FSU effective interaction has not yet been explored in any clustering studies and given its nature it seems to bear most potential for helping to understand the physics of clustering in light-to-medium mass nuclei. 

The FSU interaction describes well the $^{19}$F spectrum, there is a good correspondence, within a few hundreds of keV, between the shell model results and experimental data. While the shell model description of the states in $^{19}$F is very good, this subject
is not of the main interest in this work. Our goal here is {to complement experimental evaluations and studies reported in the previous section and }to look at the alpha clustering and overall alpha+core dynamics that is most interesting in the channels with low angular momentum. Due to the small centrifugal barrier these are the situations where decays into alpha channels are strong, and the influence of these channels in reactions is large. This is particularly relevant for  astrophysics. 
Thus, we limit consideration to the partial waves with $\ell=0$ and $\ell=1.$ Also, we concentrate on the states above the threshold. 
Alpha separation energy in $^{19}$F is at 4.013 MeV so states above that would not be visible in alpha scattering experiments. Alpha separation energy in $^{20}$Ne is 4.7299 MeV.
Detailed shell model predictions obtained using FSU interaction and using the techniques of Ref.~\cite{08}
are compared with experimental data on clustering in the following tables 
\ref{tablL3} for $\ell=0$ and in \ref{tablL4} for $\ell=1$.

The mean-field methods and the particle-hole hierarchy of excitations has been at the core of the nuclear shell model. Even the term shell model as opposed to a more generic term configuration interaction technique used elsewhere, highlights a prevalence of states to be organized by their particle-hole excitation across shells above the Fermi surface. 
{As we mentioned in the introduction, our previous theoretical efforts  \cite{06,07} that used fully mixed calculations within two oscillator shells were successful for states mostly within 0$\hbar\omega$ excitation, but some positions of states in the spectra and clustering 
collectivization were not reproduced for the negative parity (mainly 1$\hbar\omega$) states and 
for states of both parities dominated by higher cross shell excitations. 
Yet, this and multiple other experiments indicate that at the microscopic level strong clustering strength that cumulatively exceeds the single-particle 
Wigner limit comes from different alpha channels that can be identified with a different number of nodes in the relative alpha plus core wave function. This  suggests collectivization toward clustering channel within each set of states of  a given harmonic oscillator 
quanta of excitation $\hbar\omega$. 
Recently published study of $^{20}$Ne \cite{dreyfuss:2020} which produces similar results and highlights the effectiveness of the algebraic
techniques built around harmonic oscillator shell structure supports this idea. 
Thus, in this work we approach with different theoretical strategy and start with the shell model interaction that is built with the particle-
hole excitation hierarchy in mind. 
This helps us to understand the clustering collectivization and to have a 
clear harmonic oscillator based identification of clustering channels based on the number of oscillator quanta. Effects of many-body mixing and interaction with the continuum are to be explored later. }

Let us first consider the lowest $J=0$ states in $^{20}$Ne and examine them in terms of  $^{16}$O$+\alpha$ in $\ell=0$ channel,  see Tab. \ref{tablL3}. Within the harmonic oscillator picture the lowest allowed configuration involves placing the four nucleons onto $sd$ shell. If we assume that in the same basis the alpha particle has no intrinsic harmonic oscillator excitations then all 8 quanta must be carried out by the relative $^{16}$O$+\alpha$ motion which amounts to the relative wave function having $n=4$ nodes.
This number is listed in the third column of Tab. \ref{tablL3}. In our notations the total number of oscillator quanta is given by $2n+\ell,$ where $n$ is the number of nodes in the radial wave function not counting origin, details of the oscillator algebra can be found in many textbooks, see for example Ref. \cite{32}. 
The excitation energy of the first excited $0^+$ state at 6.7 MeV agrees well with experiment, this state is also clustered with spectroscopic factor SF$_\alpha$=0.14. {Here we define spectroscopic factor as an overlap 
of the normalized alpha channel wave function with the state of interest, squared, 
for details see Ref. \cite{08}. The magnitude of this SF is expected to be 
roughly proportional to the reduced width $\gamma_\alpha$ obtained from experiment as the ratio of observed decay width to the width obtained 
for a resonance at the same energy in the potential model.}
Both lowest states are $sd$ states coupling to the alpha channel wave function with $n=4.$ The next $0^{+}_3$ state predicted at 7.5 MeV  is likely a counterpart to the next known state at 7.19 MeV. This state is a $2\hbar\omega$ state dominated by the two particle-hole excitation of nucleons from $p$ to $sd$ shells. This is an $n=5$ node state with respect to the clustering channel, but in agreement with experiment this state has a much smaller alpha SF. 
The main clustering strength for alpha scattering in $n=5$ channel appears in our theoretical model at higher energy, around 13.5 MeV. In addition the shell model predicts $4\hbar\omega$ state $0^+$ state at 10 MeV ($n=6$). Configuration mixing and coupling to the continuum suggest these states as candidates for explaining a broad alpha resonance seen in experiments. It is instructive to compare these results with those reported in \cite{07}. The previous calculations were done using several older and more restrictive theoretical models, the $n=4$ channel results ($0\hbar\omega$ valence space) agree well with those from USDB hamiltonian \cite{33} restricted to $sd$ shell, the $2\hbar\omega$ states coupled to $n=5$  channel emerge from consideration of $p-sd$ space with Hamiltonian from Ref. \cite{34}. The $p-sd$ hamiltonian used in work \cite{07} allowed for $\hbar\omega$ mixing but the valence space limitation limits its applicability to $n=4$ and $n=5.$ The $4\hbar\omega$ excitations are not reasonable to discuss without the $fp$ oscillator shell. It appears that the mixing between $0\hbar\omega$ and $2\hbar\omega$ in the Hamiltonian from Ref. \cite{34} is excessive, giving a $0^+_3$ state a much larger alpha SF. 
Both, the previous work and these results do not reproduce the broad $0^+_4$ state but the emergence of the $4\hbar\omega$ state in this study which couples to the alpha channel with $n=6$ nodes offers a way to explain the appearance of significant new alpha strength coming with a new alpha channel that has $n=6$ nodes in the alpha-core relative wave function. It is likely that configuration mixing and coupling through the continuum redistribute and lower this strength, making $0^+_4$ very broad. Further theoretical efforts, larger valence space and a more elaborate models are needed to understand the lowering of the alpha strength. 

\begin{table}[htb]
\centering
\begin{tabular}{@{}|c|c|c|c|c|c|c|c|@{}}
\hline
$J_{i}^\pi$ &$E$(MeV) & $n$ & SF$_\alpha$ &$E(MeV)$  & $\gamma_\alpha$ & SF$_p$& SF$_p$\\
th &th  & th &th & exp & exp &exp & th \\
\hline
$0^{+}_{1}$ & 0 & 4 & 0.755 & 0 &  &&\\
$0^{+}_{2}$ & 6.698 & 4 & 0.143 & 6.725 & 0.47 &&\\
$0^{+}_{3}$ & 7.547 & 5 & 0.007 & 7.191 & 0.017 &&\\
$0^{+}_{4}$ & 10.121 & 6 & 0 & 8.7 & broad && \\
$0^{+}_{5}$ & 11.885 & 5 & 0.093 &  &  &&\\
$0^{+}_{6}$ & 11.908 & 4 & 0.002 &  &  &&\\
$0^{+}_{7}$ & 12.160 & 5 & 0.002 &  &  &&\\
$0^{+}_{8}$ & 13.521 & 5 & 0.246 &  &  &&\\
\hline

$1/2^{-}_{1}$ & 0.468 & 4 & 0.706 & 0.110 &  &0.24& 0\\
$1/2^{-}_{2}$ & 6.900 & 4 & 0.020 & (6.095) & &0.12& 0.04 \\
$1/2^{-}_{3}$ & 7.092 & 4 & 0.041 & 7.048$^*$ & 0.12 &&0.02\\
$1/2^{-}_{4}$ & 7.292 & 5 & 0.006 & 7.702 & && -\\
$1/2^{-}_{5}$ & 7.856 & 4 & 0.101 & 6.540$^*$ & 0.53 &&0.11\\
$1/2^{-}_{6}$ & 8.761 & 4 & 0.003 &  &  &&0.02\\
\hline
\end{tabular}
\caption{Lowest states coupled to  $\ell=0$ channel. Upper part of the table shows states in $^{20}$Ne for the  $^{16}$O$+\alpha$ reaction and lower part corresponds to $^{19}$F and $^{15}$N$+\alpha$ reaction. Columns identify state, theoretical excitation energy, number of nodes in the alpha channel, experimental energy, experimental alpha reduced width, experimental proton spectroscopic factor and theoretical proton spectroscopic factor. The labels in the second row ``th" or ``exp" refer to results coming from theory and experiment, respectively. Correspondence between data from theory and experiment is not a firm assignment, see discussion in text. The states assessed in this work are marked with $^*.$\label{tablL3}}
\end{table}

\begin{table}[htb]
\begin{tabular}{@{}|c|c|c|c|c|c|c|c|}
\hline
$J_{i}^\pi$ &$E$(MeV) & $n$ & SF$_\alpha$ &$E(MeV)$  & $\gamma_\alpha$ & SF$_p$& SF$_p$\\
th &th  & th &th & exp & exp &exp & th \\
\hline
$1^{-}_{1}$ & 6.982 & 4 & 0.381 & 5.79 & 1.4  && \\
$1^{-}_{2}$ & 7.918 & 4 & 0.379 & 8.708 &  &&   \\
$1^{-}_{3}$ & 8.957 & 4 & 0.010 & 8.854 & &&  \\
$1^{-}_{4}$ & 10.529 & 4 & 0.005 &  &  && \\
\hline
$1/2^{+}_{1}$ & 0.000 & 3 & 0.874 & 0.000 &   &0.42&0.76\\
$1/2^{+}_{2}$ & 6.060 & 4 & 0.311 & 5.333$^*$ & 1.16  && -\\
$1/2^{+}_{3}$ & 6.212 & 3 & 0.034 & 6.255 &   &0.19&0.13  \\
$1/2^{+}_{4}$ & 7.199 & 4 & 0.027 & 5.938 &  &0.014 & -\\
$1/2^{+}_{5}$ & 8.801 & 3 & 0.003 & 8.135 &   &0.156& 0.50 \\ 
\hline
$3/2^{+}_{1}$ & 1.770 & 3 & 0.672 & 1.554 &   &1.01 &0.79 \\
$3/2^{+}_{2}$ & 4.877 & 4 & 0.003 & 3.908 &   &&- \\
$3/2^{+}_{3}$ & 6.819 & 3 & 0.019 & 6.497 &   &0.133 &0.04 \\  
$3/2^{+}_{4}$ & 6.937 & 4 & 0.633 & 5.488$^*$ & 0.98  &&-\\ 
$3/2^{+}_{5}$ & 7.080 & 3 & 0.136 & 6.528 &    &&0.01\\
$3/2^{+}_{6}$ & 7.847 & 4 & 0.040 & 7.262 &   && -\\
\hline
\end{tabular}
\caption{Same as Table \ref{tablL3} but for $\ell=1$ channel.  The $1/2^{+}$ and $3/2^{+}$ spin orbit partner states that are listed separately. \label{tablL4}}
\end{table}

Let us now turn to an analogous situation in  $^{15}$N$+\alpha$ reaction. Because of the $0p_{1/2}$ proton hole in $^{15}$N  the $\ell=0$ channel with $n=4$ nodes would couple to $1/2^-$ 1$\hbar\omega$ states in $^{19}$F. 
Roughly speaking, alpha particle in this relative motion adds 8 oscillator quanta to the system by placing 4 nucleons on the $sd$ shell. See lower part of  Tab. \ref{tablL3}. The lowest  $1/2^-$ state predicted by the shell model at 0.47 MeV appears to correspond to this situation and has a large alpha SF.  The experimental counterpart at 0.11 MeV of excitation is below the alpha threshold for direct scattering. Above that, both theory and experiment have a series of $1/2^-$  states starting at about 6.5 MeV of excitation. The state seen at 6.54 MeV with reduced width of 0.53 is a likely clustering analog to $0^{+}_2$ in $^{20}$Ne. In theory this state appears at 7.8 MeV and absorbs the remaining strength for alpha in $\ell=0$ $n=4$ channel. The theoretical alpha SF's  0.14 for $^{20}$Ne and 0.1 $^{19}$F are similar.  In our model we do not consider any mixing between different $\hbar\omega$ states, of course this mixing should be present, but in nearly spherical nuclei and without other significant collective dynamics we expect this mixing to be small. The lack of mixing would suggest 
that particle decays from states with larger number of excitation quanta would be blocked. 
{
This seem to be supported by experiments, the discussed in the following text spin-orbit analog states $1/2^+$ at 5.333 MeV and $3/2^+$ at 5.488 MeV for $\ell=1$ $n=4$ channel are not seen in $^{18}\rm{O}(d,n)$ reactions \cite{terakawa:2002} although other states of the same spin and parity 
below and above in excitation energy are seen, see Tab. \ref{tablL4}
}
In our model the $1/2^-_4$ state 
that appears at 7.292 MeV of excitation is $3\hbar\omega$ state that couples to $n=5$ node alpha channel wave function. The proton decay of 
this state to the ground state is suppressed because it would require proton to carry out 5 oscillator excitation quanta and effectively 
decay from $2p_{1/2}$ orbit of the $pfh$ oscillator shell, that is very high. The selection rules related to the number of oscillator quanta are 
are helpful in discussions of other transitions. 
The $1/2^{-}_{4}$  state in $^{19}$F could be associated with $0^+_3$ in $^{20}$Ne which is of 2$\hbar\omega$ type and thus both states
would couple to $n=5$ $\ell=0$ alpha channel.  However, both states have small SF$_\alpha$ to this channel (0.007 in $^{20}$Ne and 0.0055 in $^{19}$F) so these are not cluster states. 

The observed in this work  $1/2^-$  state at 7.0 MeV has a reduced width of 0.12 and therefore is unlikely to be 3$\hbar\omega.$  The shell model predicts several other states around 7 MeV of excitation that capture enough alpha strength in the $n=4$ channel. The two states at 7.048 MeV and 6.540 observed in experiments are likely the $1/2^{-}_{3}$ and $1/2^{-}_{5}$ states that are both coupled to $n=4$ alpha channel. These states being near in the spectrum, of the same spin-parity, and having the same number of oscillator quanta obviously mix and share the alpha strength. Based on the alpha channel coupling strength we identify 6.540 MeV state with shell model one $1/2^{-}_{5}$ at 7.856 but this identification is subjective.  In stars this state can provide a path for generation of $^{19}$F via $(\alpha,\gamma)$ process  \cite{03}. Experimentally the gamma width is not known but theory predicts two main gamma decay branches: E1 to the $1/2^{+}$ ground state
width 0.14 eV (B(E1)=0.0011 W.u.) and M1 to the first excited $1/2^{-}_{1}$ state with width 0.06 eV  (B(M1)=0.012 W.u.)

It is interesting to note, that no counterpart for the  broad 0+ state at 8.7 MeV  has yet been seen in $^{19}$F 
which suggests that the structure of this state 
and its strong coupling to continuum is indeed influenced by special circumstances.

In the $\ell=1$ channel there is a broad $1^-$ state observed at 5.79 MeV in $^{20}$Ne. In theory there are two states predicted at 6.9 and 7.9 MeV that are strongly coupled and share nearly full alpha strength in $\ell=1$ $n=4$ channel. Strong coupling to a decay channel is known to cause a super-radiance mechanism in overlapping resonances leading to full decay width being absorbed by one of the states \cite{35,36}. Thus the super-radiant $1^-$ is likely the state seen in experiments and redistribution of the width that this theory is unable to describe is not surprising. 

The comparison between $^{15}$N$+\alpha$ and $^{16}$O$+\alpha$ is more interesting in $\ell=1$ because negative parity 
of the relative motion allows both $n=3$ channel $^{15}$N$+\alpha$ while this channel is Pauli blocked for $^{16}$O$+\alpha$.
Effectively a proton hole in $^{15}$N can be occupied by one of the protons from an alpha particle in $^{15}$N$+\alpha$ which is not possible in 
the case of $^{16}$O.   {Difficulty of the previously used theoretical methods to describe odd-parity alpha channels in $^{20}$Ne adds 
relevance to this comparison.}

Let us discuss the  $n=4,$ $\ell=1$ channel. The scattering of $^{15}$N$+\alpha$ in this channel would populate 2$\hbar\omega$ states in $^{19}$F. Indeed, the second excited $1/2^+_2$ state predicted at 6.06 MeV and $3/2^+_4$ predicted at 6.94 MeV both have this structure and are strongly coupled to this alpha channel. This is consistent with experiments where these states appear at 5.33MeV ($1/2^+$) and 5.49 MeV ($3/2^+$). The $0\hbar\omega$ states should have an appreciable single particle spectroscopic factor which can be measured in the $^{18}$O ($d$,$n$) reaction. The correlation between the calculations and the experimental results is evident in Table  \ref{tablL4}.

{
Our theoretical approach is certainly not not perfect, 
mixing of states and involvement of the scattering continuum using more advanced theory like continuum shell model \cite{38,39,40} is yet to be done.} However, the fact that strongly clustered $1/2^+$ and $3/2^+$ are  spin-orbit partners in the $n=4,$ $\ell=1$ channel is transparent; 
this channel and the corresponding broad $1^-$ state are well known in $^{16}$O$+\alpha$ reaction. In $^{19}$F the $1/2^+$ and $3/2^+$ clustering states in $n=4$ $\ell=1$ channel are $2\hbar\omega$ states which should suppress  their particle spectroscopic 
factors and may have an effect on their gamma decays. 
Assessing this spectroscopic information from experiment,  exploration of the channel mixing via resonating group method, and study of configuration mixing related to channel coupling and continuum are of interest.  
 
\section{Conclusions}
In this work we explore  $^{19}$F and its structure as $^{15}$N$+\alpha;$ {we determine parameters of several} resonances populated in the $^{15}$N ($\alpha$,$\alpha$) elastic scattering. The  $^{19}$F plays an important role in astrophysics and its structure is central for the development of theoretical understanding of the nuclear many-body problem. As compared to oxygen chain, an extra proton in fluorine isotopes makes a huge structural difference changing the mean-field shape, pairing properties, and extending the neutron drip line much further in the mass-number \cite{40}. 

This work also pursues the goal of exploring the role of an extra proton degree of freedom in clustering  properties, and especially in comparison between $^{15}$N$+\alpha$ and $^{16}$O$+\alpha$ reactions and correspondingly alpha structure of states in $^{19}$F and $^{20}$Ne. In this work we concentrate on the channels with relative motion in the lowest partial waves with $\ell=0$ and $\ell=1$ which couple to the low-lying states and due to the small centrifugal barrier are most impactful in structure-reactions physics and in astrophysics. 
We find that the clustering structure prevails; for $\ell=0$ we identify a 6.540 MeV $1/2^-$ state that appears to be a counterpart of 6.725 MeV state in $^{20}$Ne with alpha moving relative to the core in a state with $n=4$ nodes in the radial wave function. The situation with $\ell=1$ is interesting, here in $^{20}$Ne the $1^{-}$ alpha strength that appears in 5.79 MeV state comes in the scattering channel $^{16}$O$+\alpha$ with $n=4$ nodes; in $^{19}$F lowest states are coupled to a different $n=3$ channel which is not blocked by the Pauli principle, and yet 
in this work we were able to identify cluster resonant states in $^{19}$F  representing  $^{15}$N$+\alpha$ relative motion with $n=4.$
The states in $^{19}$F,  $1/2^+$ at  5.333 MeV and $3/2^+$ at 5.488 MeV, are spin orbit partners coupling $1/2^{-}$ ground state of  $^{15}$N
with orbital $\ell=1$ motion of alpha. 

In this work we were able to make a substantial progress in understanding of clustering from a theoretical perspective, we take advantage of a 
new phenomenological shell model Hamiltonian \cite{29,30,31} that has been developed to study 
cross-shell particle hole excitations.  While particle-hole hierarchy in the theoretical approach may seem like a disadvantage in this work it played 
a crucial role in identifying clustering channels, allowing to determine origins of seemingly excessive clustering strength observed in experiments. 
In particular, a clear separation between scattering states with different number of radial nodes allows to cleanly establish spin-orbit partner 
states in $^{15}$N$+\alpha,$ $\ell=1$ channel, while accounting for all other resonances and their strengths in $^{19}$F.  
This resolves many issues encountered in previous works \cite{06,07,41}. 
{Prevalence of clustering and the emergence of strongly clustered states from a microscopical perspective appears to 
represent collectivization of states with a certain number of oscillator cross shell excitations. The reasons for this collectivization, its enhanced 
strength near thresholds,  and apparent lack of mixing of states with different particle-hole nature are yet to be studied.}
The particle-hole hierarchy also suggest suppression of particle and electromagnetic transitions and offers avenues for  experimental assessment of channel mixing and evaluation of continuum effects. This suppression may play an important role in astrophysical process and should be considered when going beyond a purely statistical treatment of reactions. {For the first time we were able to 
discuss the spin-orbit interaction for clusters from a microscopic perspective and compare it with observations; this interaction appears to be very weak and due to many-body complexity it is impossible to separate any systematic strength that is not consistent with zero.}

\begin{acknowledgments}
This material is
based upon work supported by the U.S. Department of Energy Office of Science, Office of
Nuclear Physics under Award Number DE-SC0009883.
Authors also acknowledge SSH2020014 project funded by Nazarbayev University, the Ministry of Education and Science of the Republic of Kazakhstan [state-targeted program number BR05236454] and [young scientists’ research grant number AP08052268].

\end{acknowledgments}

\bibliographystyle{myprc}

\end{document}